# High Performance Computer Acoustic Data Accelerator: A New System for Exploring Marine Mammal Acoustics for Big Data Applications

*Peter Dugan[1], John Zollweg[1], Marian Popescu[1], Denise Risch[3], Herve Glotin[4]*  
*Yann LeCun[2] and Christopher Clark[1]*

(1) Bioacoustics Research Program (BRP), Cornell University, Ithaca, NY, USA  
(2) The Courant Institute of Mathematical Sciences, New York University, NYC, USA  
(3) Northeast Fisheries Science Center, NOAA, Woods Hole Oceanographic Institute, MA, USA  
(4) Institut Universitaire de France, CNRS LSIS and USTV, FR

*Abstract*: **This paper presents a new software model designed for distributed sonic signal detection runtime using machine learning algorithms called *DeLMA*. A new algorithm--Acoustic Data-mining Accelerator (ADA)--is also presented. ADA is a robust yet scalable solution for efficiently processing big sound archives using distributing computing technologies. Together, DeLMA and the ADA algorithm provide a powerful tool currently being used by the Bioacoustics Research Program (BRP) at the Cornell Lab of Ornithology, Cornell University. This paper provides a high level technical overview of the system, and discusses various aspects of the design. Basic runtime performance and project summary are presented. The DeLMA-ADA baseline performance comparing desktop serial configuration to a 64 core distributed HPC system shows as much as a 44 times faster increase in runtime execution. Performance tests using 48 cores on the HPC shows a 9x to 12x efficiency over a 4 core desktop solution. Project summary results for 19 east coast deployments show that the DeLMA-ADA solution has processed over three million channel hours of sound to date.**

I. INTRODUCTION

Nearly every branch of science is experiencing an explosion in the amount of data collected and available for analysis. From *in situ* sensor networks to remote sensing satellites, enormous stores of ocean data are being amassed from a plurality of sources (e.g., see marinexplore.com). This includes acoustic sensors that are the mechanism by which passive acoustic data is acquired.

As in many big data fields, the main challenge for acoustic monitoring is in the processing and analyzing the vast amounts of collected data. In the past two decades, the bioacoustic sciences have made significant advances in software for collecting and analyzing both archived and real-time systems [1-3]. Despite these advances, large amounts of acoustic data remain unprocessed. There are many challenges, including size of the data banks (i.e., terabytes and beyond), and the pervasive lack of standardization, resources, and systems. And while there are systems capable of automatically processing big data archives, they have been not readily available to the scientific community. In an attempt to address big data concerns in Bioacoustics, we have designed and constructed a scalable, high performance computing (HPC) system for processing large stores of acoustic data.

Data mining algorithms can be combined with HPC systems and used for a variety of tasks associated with time series data, such as compression and acoustic modeling. The system described herein uses image processing techniques on 2D time-frequency spectrogram arrays to detect and classify vocal patterns. The exact data mining algorithms used in this work are not new to this research; however references are provided for additional details.

The system approach section presents a high level understanding for HPC computing as it relates specifically to the needs for hosting large sound archives at the Bioacoustics Research Program. The basic concepts for serial and distributed processing models are presented along with three primary requirements needed for an efficient scalable design. The design section describes process flows for the DeLMA runtime and the ADA algorithm. An interface description along with a process flow for incorporating data mining algorithms is described at a high level. Performance measures were conducted for two groups of experiments. The first was intended to provide a baseline between older serial systems, to then compare to newly distributed configurations. This first example uses a desktop for the serial system, and a 64 core distributed computer for the HPC system. The second experiment was designed to show how a current analyst computer, using a multi core desktop, might compare to a distributed HPC machine. Last, project results for 19 deployments were processed using a variety of data mining algorithms, with performance results at a project level shown and discussed in relation to the DeLMA software.

## II. SYSTEM APPROACH

### A. Distribution Model

Figure 1 illustrates the two processing models, serial and distributed. Both have four main resource components: data, algorithms, runtime and processing computer. *Data* represents both input and output formats stored using a time series sequence. Inputs are sound archives typically audio formats, used to support single or multiple channels. Data mining algorithms, or routines, are specifically designed to find vocal sounds produced by whales or other sources, and systematically interface through a software runtime. In traditional systems, the process is done using a single computer which entails reading the sound data, executing data mining code and producing detection results. If all steps are done in sequence, then the process is considered serial. Figure 1(a) shows a systems diagram illustrating this process.

The distributed processing model is shown in Figure 1(b). In comparison to the serial case, a single computer is replaced by multiple systems running several cores. Aside from the computer hardware, the fundamental difference between the serial and distributed models is the software runtime. Replacing the serial runtime software with an HPC version allows for multiple computers to gain access to the sound data, referred to as distributed processing.

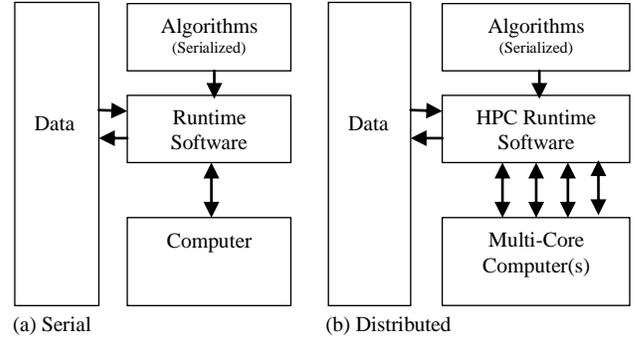

Figure 1. System for data mining sound archives, components consisting of *Data*, *Algorithms*. (a) Serial system uses *runtime software* and standard *computer*. (b) HPC System uses single or distributed *multi-computer(s)* and *HPC runtime software*.

### B. System Requirements

Three main requirements were established for this work, each summarized in Table 1. First, the same HPC runtime could be used across serial or distributed systems, that is, runtime behavior would be independent of the distribution model. The goal for requirement one was to offer an interface layer between the user and the system that abstracts the complex interconnection of hardware and software. While the abstraction is convenient for the user, it does not mean that all physical hardware configurations will perform the same. Slow networks or unbalanced computer resources are transparent to the abstraction and may impact runtime performance. The second requirement was to develop a relatively easy method to interface data mining algorithms readily available through the open source community. This requirement would hopefully provide more options for solving problems and enhancing collaboration. The goal was to provide a mechanism to interface algorithms to the HPC runtime. Keeping the algorithms serial helps to provide stable and consistent data mining software results. The third requirement was that equal portions of work be distributed among the resources, creating execution across all cores of the distributed system. Uniform workflow would allow for easier scalability.

## III. SYSTEM DESIGN

### A. DeLMA Runtime

Initial phases of this work focused on developing cost effective hardware and software components capable of processing large datasets or sound deployments. The DeLMA runtime design is best shown using the system flow diagram in Figure 2. Step 1, *Initialization* contains instructions for initializing, and learning the system resources. This includes initial creation of the data mining algorithms and establishing the default thresholds. An analysis of system processing capability is also conducted at initialization where the number of CPU cores, memory and data locations is loaded. Step 2, *Setup*, uses the information that was created during initialization, wherein the options are presented to the user. Job *Setup* allows the user to select several options and create jobs for execution. This includes configuring the number of CPU's to allocate for a specific job. In many cases users can adjust the processing speed by selecting more or fewer cores to use based on the size of the sound datasets, the number of independent channels and the list of data mining algorithms used in the analysis. Once the job has been established, the Step 3 *ADA* algorithm studies the user selection and divides up the sound archive into data blocks. Data blocks and user setup information are then sent at Step 4, *Processing*, to the processing resources as a single process using multiple data streams, or spmd [10]. *Output* occurs in Step 5.

| Requirement | Description |
| --- | --- |
| Req-1 | The same HPC runtime will be used across serial or parallel-distributed configurations; runtime behavior is independent of the distribution model. |
| Req-2 | Data mining algorithm execution will be independent of the behavior of the distribution model; rather the runtime will be responsible for distributing workload. |
| Req-3 | Equal portions of work will be distributed among the resources creating execution across all cores of parallel-distributed system. |

Table 1. Main design requirements for the HPC DeLMA runtime software.

Since ADA divides the work using a uniform work strategy, all the resources finish at approximately the same time, allowing data mining output results from the algorithms to be gathered using Step 5.

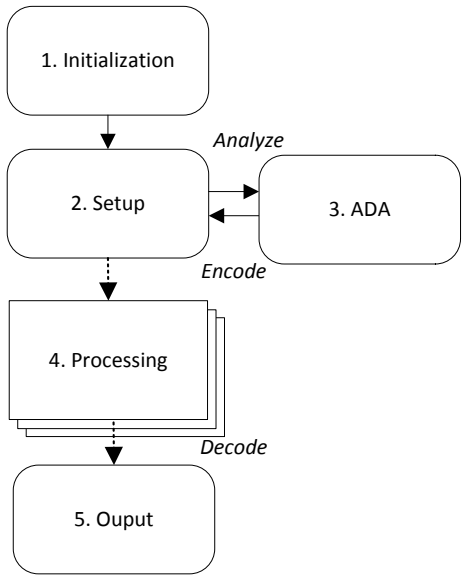

Figure 2. System flow diagram illustrating design flow for the DeLMA runtime and algorithm interface.

### B. ADA Algorithm

As depicted at Figure 2 (the system flow diagram), the acoustic data acceleration (ADA) algorithm is Step 3. According *Req-3*, Table 1, ADA divides the work using a uniform work strategy. This is done by analyzing the setup parameters established at Step 2. Once analyzed, a vector is encoded by using a series of start and end pairs that describe each data-block. The encoding process accounts for gaps in the data, which may be a result of duty cycling the sensors or errors due to problems in the sound archive.

Routines use the encoding vector to distribute the data array to all the resources, where each core will have an independent block of data to process (see codistributed commands) [10]. Data mining algorithms are then distributed to each core and begin execution on the data blocks. Since each core performs equal portions of work, tasks finish at approximately the same time, ready to assemble the outputs from the various data blocks. DeLMA Output, Step 5, decodes the distributed arrays by inverting the distribution map (see gathering) [10].

### C. Data Mining, Algorithm Interface

Rules for creating the data mining interface meet Table 1 Requirements, *Req-1* and *Req-2*. The top two requirements are satisfied by providing an object-oriented wrapper that coincides with Figure 2, Steps 1, 2, 4 and 5, and imply that the algorithm can be decomposed in its original form to satisfy parameter *Initialization,* user *Setup*, code *Processing* and result *Output*. Step 4 is responsible for divvying up the sound data and running the data mining algorithms on each section. In other words, whether DeLMA uses serial or distributed mode, the data mining routines remain serialized throughout the processing sections of the sound archive, thereby satisfying Req-3 as well.

## IV. SETUP

### A. Data Mining Algorithms

A collection of data mining algorithms were selected for the extraction of sound events. Description of the data mining algorithms is beyond the scope of this work, however references are provided in Table 2 for the reader. Each algorithm in Table 2 was converted and added to the DeLMA runtime using the steps described in Figure 2. Each data mining algorithm shown in Table 2 has an ID, a signal type and a description. The ID is a unique field used to identify the algorithm within the context of this paper. The algorithm type provides the species and signal shape. For simplicity, two different signal shapes are used as descriptors, a "sweep" or "pulse." A sweep is a modulated call type which varies in time and frequency. A pulse is a single, high energy signal having short time duration and relatively high bandwidth. Algorithms ID = [4-7, 10-12] use a series of single pulse shapes, or pulses, to describe the vocalization, these are labeled accordingly and ID = [1-3, 8, 9] are sweeps.

| Algorithm ID | Species (Signal Type) | Algorithm description and reference. |
|---|---|---|
| 1 | Right Whale (sweep) | Custom multi-stage detection-recognition algorithm *isRAT*. [11-14]. |
| 2 | Right Whale (sweep) | Detection-classification using histogram of oriented gradients [15-17]. |
| 3 | Elephant (pulse) | |
| 4 | Seismic air gun (pulses) | Multi-stage energy detection, using connected region analysis [18-21]. |
| 5, 6, 7 | Sperm, Minke, Fin (pulses) | |
| 8 | Right Whale (sweep) | |
| 9 | Brydes Whale (sweep) | Data-template and matched filtering concepts [22, 23]. |
| 10 | Fin Whale (pulses) | |
| 11 | Minke Whale | Multi-stage energy detection, connected region analysis and pulse-train, cross correlation [19, 21, 24, 25]. |
| 12 | Fin Whale (pulses) | |

Table 2. Data mining algorithms currently used in the DeLMA runtime.

*B. Processing Methods*

Three different computer configurations were used to compare data processing performance; these are shown in Table 3. *Method One* is a serial configuration created with a desktop computer running a single core. *Method Two* was the same hardware as Method One except the DeLMA software used 4 cores during the runtime. *Method Three* used an HPC machine running the DeLMA software as distributed configuration, using a maximum of 64 cores. Referring to Table 3, *Methods One and Two*, the desktop computer workstation was an Intel Xeon E5-2620 @ 2.0 GHz. Desktop configuration represents the typical system used by analysts. Computer for *Method Three* was an HPC machine which consisted of 4 system boards, each having dual-quad core Intel Xeon E-2670 @ 2.6 GHz. The HPC configuration represented a network type system, similar to a cloud based application.

*A. Experiments*

Runtime performance was measured using two experiments. Experiment one compared performance between Method One and Method Three, and was intended to test the older processing models used by analysts, or serial operation with a new HPC distributed configuration. The HPC configuration was setup using all 64 available cores. A dataset consisting of 4.38 TB of 2 kHz sounds spanning over 172,896 hours was staged on the desktop computer. The same dataset was staged on the HPC machine and run using the DeLMA software with 64 cores. Both methods used a standard data mining algorithm as a benchmark for processing.

The second experiment was designed to compare two different distributed computer systems, Method Two and Method Three configurations, Table 3. The desktop computer used 4 cores and the HPC system used 48 cores. This experiment considered different sizes and sample rates for the sound archive. A total of three sound archives were chosen and performance was measured for each. Archive One was 16 kHz, 592 GB spanning 5,520 hours, Archive Two was 2 kHz, 11 GB, spanning 168 hours, and three was 2 kHz, 380 GB spanning 29,808 hours. Efficiency factors were computed for both experiments by taking the ratio of total runtime (in seconds) between the HPC system and desktop systems.

Table 4 illustrates information gathered from various projects that used the DeLMA runtime between 2011-2013. A summary for each program was captured for historical purpose and metrics were collected to estimate the total number of deployments, total channel hours, average number of jobs run per project and the total channel hours processed. Table 4 also contains information about various algorithms that were used for each project. In many cases, multiple algorithms were used for a single species (e.g., right whale). DeLMA runtime was able to execute all algorithms for a single job. Project data was run multiple times to account for errors or to refine algorithm accuracy. A baseline summary for quantifying data processing metrics for historical purpose is discussed in the next section.

| | Method One | Method Two | Method Three |
|---|---|---|---|
| **Resource** | Desktop Computer | Desktop Computer | HPC Computer |
| **Computer Nodes** | 1 core | 4 core | 64 core |
| **Processor** | Intel Xeon E5-2620 @ 2.0 GHz | Intel Xeon E5-2620 @ 2.0 GHz | Intel Xeon E-2670 @ 2.6 GHz |

Table 3. Hardware methods used in this work, each using the same DeLMA runtime.

| Deployment | Channel Hours (Est.) | Job Runs | Algorithm Signal Type [ID] |
|---|---|---|---|
| Boston Harbor Excellerate | 832k | 1 | Right Whale [1] Fin Whale [7] |
| Gulf of Mexico | 350k | 3 | Sperm Whale [5] Brydes Whale [9] |
| Greenland | 5.5k | 5 | Seismic Air Gun [4] |
| Mass CEC | 25k | 3 | Minke Whale [6, 11] Right Whale [1, 2, 8] |
| Gulf of Maine | 26.3k | 2 | Minke Whale [11] Right Whale [1, 2, 8] |
| Cape Cod Bay | 21.6k | 6 | Right Whale [1, 2, 8] Minke Whale[11] Fin Whale [7] |
| Stellwagen Bank National Marine Sanctuary | 60.4k | 10 | Right Whale [1, 2, 8] Minke Whale [6, 11] Fin Whale [7] |
| Virginia | 23.5k | 2 | Right Whale [1, 2, 8] Minke Whale [6, 11] Fin Whale [7] |
| NAVFAC (32 kHz) | 10k | 2 | Sperm (PT) Right Whale [1, 2, 8] Minke Whale [11] Fin Whale [7, 12] |

Table 4. Select projects that used HPC system and DeLMA runtime software.

## V. RESULTS

### A. Serial Baseline vs. Distributed

Table 5 illustrates the runtime performance, comparing serial and distributed processing. Serial processing for the 4.38 TB data was slow using *Method One*. The job was stopped after 10% of the data was processed and took slightly over 52 hours. The complete runtime performance was extrapolated using the partial run as shown in Table 5. The job for *Method Three* finished in ample time, slightly longer than 12 hours and the runtime efficiency was computed as 44:1, see Table 5.

| Dataset | | Method Three | | Method One | | Run Time Efficiency. |
|---|---|---|---|---|---|---|
| Sample Rate | Total Hours (Size bytes) | Number of Cores | Runtime (HH:MM:SS) | Number of Cores | Runtime (HH:MM:SS) | |
| 2 kHz | 172,896 (4.38 TB) | 64 | 12:01:00 | 1 | 528:00:00 | **x44** |

Table 5. Results desktop (serial) versus HPC-DeLMA (distributed) performance, *Method One* vs. *Method Three*.

### B. Distributed Desktop vs. HPC

Table 6 summarizes the runtime performance comparing the distributed mode between the desktop computer and the HPC system. Three data scenarios were used for the comparison. The first experiment, involving 16 kHz sound data spanning 5,520 hours of sounds took 162 hours to process using a 4 core desktop computer. The HPC utilizing 48 cores was completed in just over 12 hours and 46 minutes and an overall efficiency of 13:1 was realized after comparing both systems. The second experiment shown in Table 6 was a 2 kHz data set spanning only 168 hours. In comparison, the HPC using 46 cores performed at roughly 29 minutes and the desktop using 4 cores took 4 hours and 53 minutes with an overall efficiency of 10:1.

| Dataset | | Method Three (HPC) | | Method Two (Desktop Server) | | Run Time Efficiency. |
|---|---|---|---|---|---|---|
| Sample Rate | Total Hours (Size bytes) | Number of Cores | Runtime (HH:MM:SS) | Number of Cores | Runtime (HH:MM:SS) | |
| 16 kHz | 5,520 (592 GB) | 48 | 12:46:40 | 4 | 162:00:00 | **x13** |
| 2 kHz | 168 (11 GB) | 48 | 00:29:10 | 4 | 04:53:00 | **x10** |
| 2 kHz | 29,808 (380 GB) | 48 | 03:57:08 | 4 | 36:00:00 | **x9** |

Table 6. Results showing distributed desktop and HPC performance comparison.

The last data set in Table 6 was a 2 kHz archive spanning 29,808 hours. The HPC system processed the data in 3 hours and 57 minutes, while the desktop computer took 36 hours. The runtime efficiency was 9:1.

*C. 2011-2013 Projects*

Table 7 illustrates a summary of Table 4 projects studied using the DeLMA runtime between 2011 and 2013. For the 19 selected deployments, a total of 1.44 million channel hours of sounds were used. An average of over 5 jobs per project, totaling 3.6 million channel hours of sounds, were processed for the combined study.

| Sample of East Coast Deployments Processed Between 2011-2013 | Channel Hours (million hours) | Average Number (Jobs Run / Project) | Total Channel Hours Run (million hours) |
|---|---|---|---|
| 19 | 1.4 | ~5.0 | 3.6 |

Table 7. Summary of East Coast data, processed through the DeLMA runtime between 2011 and 2013.

V. DISCUSSION

BRP has developed a software runtime called DeLMA. When DeLMA is combined with the ADA algorithm the system is capable of providing distributed computing solution for data mining sound archives. The design was based on requirements that provide scalability, ease of algorithm integration, and balanced processing execution. The system was field tested on several different East Coast projects collected between 2011- 2013. For this work various standard detection classification algorithms were interfaced to the DeLMA runtime for analysis.

Comparison results between a desktop server and the HPC hardware shows an estimated throughput rate of 44 times faster using 64 distributed cores over a standard analyst-grade server, running a single core. The second experiment used distributed computing to compare the workstation to the HPC system. Using DeLMA running 4 cores on the desktop workstation, efficiency ranged from 13:1 to 9:1 times faster for the 48 core setup with the HPC machine. Feasibility for utilizing the system for big data applications was tested using 19 deployments which contained a mixture of data formats. Example sets spanned 1.44 million channel hours of acoustic recordings taken from the BRP archive collection, focusing on the eastern coastal region of the United States. Fast processing provided the capability to interactively develop and rerun jobs. The example datasets were processed, on average, roughly 5 times for each deployment, resulting in a total of 3.36 million channel hours of data for the 19 project deployments.

Several key factors should be noted. First, there is a small difference in efficiency ranging from 9x to 10x between 2 kHz experiments for Methods Two and Three. However, increase sample rate to 16 kHz resulted in 13x efficiency. This suggests that the data resolution has a significant impact on computing performance and higher sample rates may be better suited for distributed HPC systems than desktop computers. As for the serial case, running 172k channel hours of sounds was not practical and the run was interrupted at 10% of the way through. However, a 64 core HPC machine was able to process the entire set in 12 hours. Lastly, comparing the 2011-2013 BRP projects, many of the deployments were run several times. Factoring these points suggests that using the DeLMA runtime may be a necessary component to studying large sound archives for future work.

This work demonstrates how large sound archives can efficiently be processed using HPC technology. For all runtime configurations, the ADA algorithm provided a scalable interface to the data, allowing DeLMA to manage input and output operations. Despite performance differences between computer architectures, the same *DeLMA* software was used in either serial or distributed configurations, allowing the research program to adjust processing to meet the needs of the project.

VI. ACKNOWLEDGMENT

Work made possible by National Oceanic Partnership Program (NOPP), Office of Naval Research (ONR) N000141210585, National Fish and Wildlife Foundation (NFWF) 0309.07.28515. Thanks to BRP research analyst and staff. Special thanks for those who show continued collaboration and support for DCL big data applications for bioacoustics, namely Dr. Sofie Van Parijs, Northeast Fisheries Science Center, NOAA, New York University, Dr. Joan Bruna, and Ross Goroshin.